\documentclass{WileyMSP-template}
\usepackage{bbm}
\usepackage{mathrsfs}
\usepackage{amsmath}
\usepackage{amsfonts}
\usepackage[colorlinks=true,citecolor=blue,anchorcolor=blue]{hyperref}
\usepackage{graphicx,epstopdf}
\usepackage{subfigure}
\usepackage{epsfig}
\usepackage{dcolumn}
\usepackage{bm}
\usepackage{color}
\usepackage[numbers,sort&compress]{natbib}
\usepackage{amssymb}
\usepackage{xcolor}
\usepackage{braket}
\usepackage{soul}
\usepackage{CJKutf8}
\usepackage{ragged2e}
\def\zuomw{\textcolor{blue}}

\begin{document}

\pagestyle{fancy}
\rhead{\includegraphics[width=2.5cm]{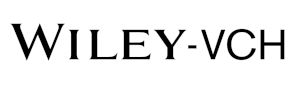}}

\title{Diagnosing Floquet Chern and anomalous topological insulators based on Bloch oscillations}

\maketitle


\author{Maowu Zuo}
\author{Yongguan Ke*}
\author{Zhoutao Lei}
\author{Chaohong Lee*}


\begin{affiliations}
Maowu Zuo, Zhoutao Lei\\
Laboratory of Quantum Engineering and Quantum Metrology, School of Physics and Astronomy, Sun Yat-Sen University (Zhuhai Campus), Zhuhai 519082, China\\
\medskip
Yongguan Ke, Chaohong Lee\\
Institute of Quantum Precision Measurement, State Key Laboratory of Radio Frequency Heterogeneous Integration, College of Physics and Optoelectronic Engineering, Shenzhen University, Shenzhen 518060, China\\
Quantum Science Center of Guangdong-Hong Kong-Macao Greater Bay Area (Guangdong), Shenzhen 518045, China\\
Email: keyg@szu.edu.cn, chleecn@szu.edu.cn
\end{affiliations}



\begin{abstract}

It is challenging to distinguish Floquet Chern insulator (FCI) and Floquet anomalous topological insulator (FATI) because of their common features of chiral edge states and far away from equilibrium.
A hybrid straight-curved waveguide array is proposed to enable topological phase transitions from FCI to FATI and show how to diagnose the two phases using Bloch oscillations.
As a proof of principle, the hybrid straight-curved waveguide array is designed as a straight honeycomb waveguide array nested in an asynchronous curved Kagome waveguide array.
Under a two-dimensional (2D) tilted potential created by the spatial gradient of refractive indices, an initial Gaussian-like wavepacket undergoes 2D Bloch oscillations, displaying quasi-quantized displacement in the FCI and no drift in the FATI.
This approach offers a direct and unambiguous method to diagnose Floquet topological phases from the bulk response.

\end{abstract}


\section{Introduction}
Floquet topological insulator,\textsuperscript{\cite{kitagawa2010topological,fang2012realizing,cayssol2013floquet,titum2015disorder,roy2017periodic,oka2019floquet,peng2020floquet,rudner2020band}} featured by topological gapped states far from equilibrium, are of great interest and versatility. 
Photonic lattices have emerged as an excellent platform for exploring Floquet topological insulators,\textsuperscript{\cite{price2022roadmap,lan2022brief,jalali2023topological,wang2023floquet}} such as Floquet Chern insulators (FCIs) in synchronous curved waveguides\textsuperscript{\cite{rechtsman_2013_photonic,lumer2013self,zhu2018topological,yang2020photonic,biesenthal2022fractal}} and Floquet anomalous topological insulators (FATIs) in asynchronous curved waveguides.\textsuperscript{\cite{ke2016topological,leykam2016anomalous,maczewsky2017observation,cheng2019observation,mukherjee2020observation,mukherjee2021observation,maczewsky2020nonlinearity,mukherjee2023period}}
Although both FCIs and FATIs support chiral edge states (CESs), they are characterized by different topological invariants: (i) non-zero Chern number of Floquet bands in FCIs and (ii) non-zero winding number of Floquet band gaps in FATIs even with zero Chern numbers for all Floquet bands.\textsuperscript{\cite{rudner2013anomalous}}
Surprisingly, it is possible to realize both FCIs and FATIs under the same driving scheme, although such systems remain rare.\textsuperscript{\cite{kitagawa2010topological,leykam2016anomalous,zhu2021symmetry}}

\indent One cannot distinguish FCIs and FATIs solely based on CESs.
Because of nonzero vs. zero Chern numbers, one may diagnose these topological phases by observing quantization phenomena related to the bulk Chern number.
Many methods have been developed to measure Chern numbers in static systems,\textsuperscript{\cite{rudner2013anomalous,nathan2015topological,zhang2020unified,jia2021high,shi2022topological}} such as quantized Hall conductance in response to static driving, circular dichroism in response to periodic driving,\textsuperscript{\cite{tran2017probing,asteria2019measuring}} and quenched dynamics (linked number, band inversion surface, and dynamic winding number).\textsuperscript{\cite{wang2017scheme,sun2018uncover,zhang2018dynamical,zhang2019dynamical,tarnowski2019measuring,zhu2020dynamic,li2021direct,jia2023unified}}
Because the Chern number is the integral of Berry curvatures over the Brillouin zone, uniform band occupation or sweeping the whole Brillouin zone is necessary.
By applying a tilted potential, even without uniform band occupation, Bloch oscillations can uniformly sample Berry curvatures in momentum space and lead to nearly perfect quantized displacements.\textsuperscript{\cite{ke2020topological,zhu2021uncovering}}
Although motivated, it is not straightforward to directly generalize these methods to the Floquet topological insulators, because of the distance from equilibrium in Floquet systems. 
Detecting FCI and FATI is highly appealing and challenging. 
Notably, the topological invariants for FATIs is defined over the full driving-period dynamics rather than stroboscopic dynamics.\textsuperscript{\cite{rudner2013anomalous}} This characteristic makes the extraction of the topological features of FATIs more challenging compared to that of some other topological phases.

In this paper, we construct a new type of Floquet models of topological phase transitions between FCI and FATI, and suggest using 2D Bloch oscillations to diagnose the two topological phases.
Our system comprises straight honeycomb waveguide arrays embedded within curved Kagome waveguide arrays, termed the Floquet honeycomb-Kagome photonic lattice; see \textbf{Figure}~\ref{f1}.
Unlike existing systems that achieve similar phase transitions,\textsuperscript{\cite{leykam2016edge}}, our scheme requires only one-dimensional (1D) periodic curving, significantly reducing the complexity of the Floquet driving.
In addition, the spatial gradient of the refractive indices in two dimensions can be introduced, which plays the role of an external force.
The periods of Bloch oscillations in two dimensions are coprime with respect to the driving period.
By preparing the initial state as a Gaussian-like momentum state in a certain non-trivial Floquet-Bloch band, we can observe near quantized displacement related to the Chern number in an overall period.
The direction of displacement is orthogonal to the effective external force.
We can then diagnose FCI and FATI via nonzero and zero quasi-quantized displacements in 2D Bloch oscillations, respectively.
Our work provides a cornerstone for faithfully realizing and identifying topological phase transitions between FCI and FATI.

\section{Model}
\begin{figure*}[!htp]
\begin{center}
  \includegraphics[width=0.5\linewidth]{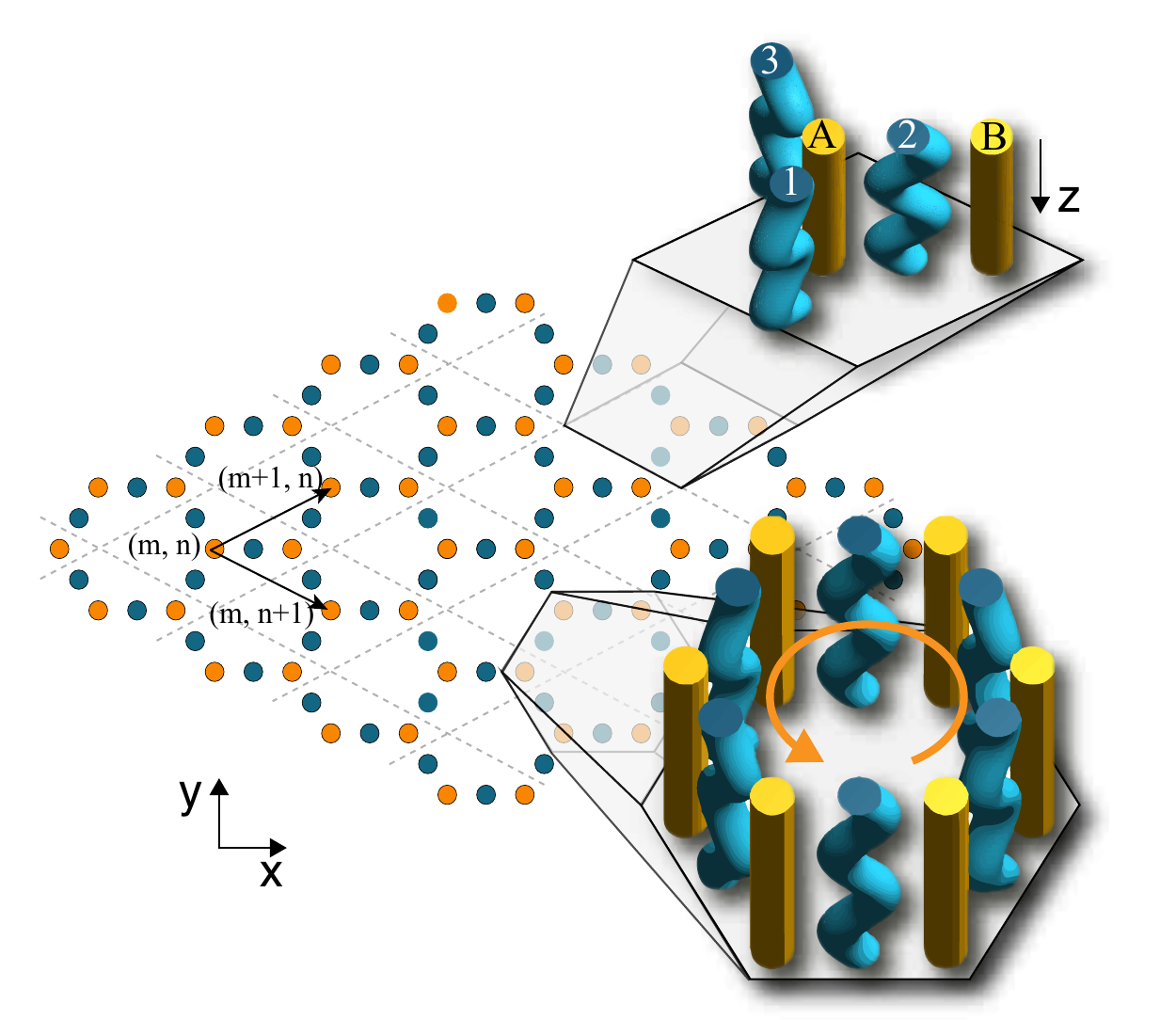}
  \caption{Schematics of Floquet honeycomb-Kagome photonic lattice. The orange dots and blue dots mark the honeycomb sublattice and the Kagome sublattice, respectively. Upper and lower insets show a unit cell and a ring of hexagonal lattice, respectively.}
  \label{f1}
\end{center}
\end{figure*}
Unlike the nested photonic lattice with 3D curving or modulated on-site refractive indices,\textsuperscript{\cite{pyrialakos2022bimorphic}} we consider a photonic lattice formed by a set of straight waveguide arrays nested in another set of asynchronous curved waveguide arrays.
We impose that each curved waveguide only changes its direction in the plane of the two sandwiched straight waveguides.
An individual curved waveguide plays the role of a carrier of light between two straight waveguides.
In addition, this structure also differs from the pure asynchronous helical of waveguide arrays,\textsuperscript{\cite{leykam2016anomalous}}
where the positions of individual waveguides change in three dimensions.
In principle, our system only involves curved waveguides in two dimensions, simpler than helical\textsuperscript{\cite{rechtsman_2013_photonic,leykam2016anomalous,yang2020photonic}} and evanescently coupled\textsuperscript{\cite{maczewsky2017observation,maczewsky2020nonlinearity,mukherjee2023period}} waveguides in three dimensions.
The asynchronous change of waveguide position not only breaks the time-reversal symmetry but also provides a chance to realize FATIs.

In our Floquet honeycomb-Kagome photonic lattice, each unit cell $\{m,n\} $ consists of two honeycomb sublattices (labeled by $\Lambda=A, B$)  and three Kagome sublattices (labeled by $j=1,2,3$), indicated by the rhombus and its zoomed-in view in Figure~\ref{f1}. 
The positions of three curved waveguides within a unit cell are described by $\mathbf{r}^{\{m,n\}}_j=\mathbf{r}^{\{m,n\}}_A+\big(d/2+R\sin(\Omega z+ \phi_j)\big)\cdot\left[\cos(\phi_j),\sin(\phi_j)\right]$, where $\phi_j=2\pi(j-2)/3$ are the initial phases, $\mathbf{r}^{\{m,n\}}_A$ is the position of $A$ waveguide, $d$ is the distance between $A$ and $B$ waveguides, $\Omega=2\pi/T$ is the driving frequency with $T$ being driving period, and $R$ is the driving amplitude. 
The lattice constant is $a=\sqrt{3}d$. In this design, the curved waveguides are asynchronously coupled to the straight waveguides with periodically-modulated coupling strength. 
The instantaneous strong coupling evolves in a loop-like manner within a ring of hexagonal lattice in Figure~\ref{f1}, effectively breaking time-reversal symmetry and enabling chiral topological transport.

To describe the system in the momentum space (see \zuomw{Supporting Information}), we derive a $z$-dependent tight-binding Hamiltonian with nearest-neighboring couplings,
\begin{equation}\label{eq2}
H(\mathbf{k},z) = - \sum_{\Lambda=A,B}\sum_{j=1,2,3}c_{\Lambda,j}(z)\psi_{\Lambda,\mathbf{k}}^\dagger\psi_{j,\mathbf{k}}\mathrm{e}^{i\mathbf{k}\cdot\boldsymbol{\delta}_{\Lambda,j}}+
 \mathrm{h.c.},
\end{equation}
with $\psi_{\Lambda(j),\mathbf{k}}^\dagger$ and $\psi_{\Lambda(j),\mathbf{k}}$ creating and annihilating light field of Gaussian modes in the $\Lambda(j)$ honeycomb (Kagome) sublattices with quasi-momentum $\mathbf{k}$, respectively. 
Here, $ \boldsymbol{\delta}_{\Lambda,j}$ is the vector displacement between the $\Lambda$ and $j$ sites. 
%
We have neglected the effective detuning of the curved waveguides which is much smaller compared to the coupling strength (see \zuomw{Supporting Information}). 
Because the spacing between the $A(B)$ honeycomb waveguide and the $j$ Kagome waveguide is periodically modulated as $s_{A(B),j}(z)=d/2\pm R\sin(\Omega z+\phi_j)$, the coupling strength $c_{\Lambda,j}(z)$ decays exponentially with the spacing $\alpha\exp(-\gamma s_{\Lambda,j}(z))$, and is also periodically modulated.
The parameters $\alpha=0.016/\mathrm{\mu m}$ and $\gamma=0.345/\mathrm{\mu m}$ are calibrated by comparing light propagation in the tight-binding model and the continuous model of a coupler composed of straight and curved waveguides (see \zuomw{Supporting Information}).

\section{Results}
\subsection{Topological phase transitions}

\begin{figure*}[!htp]
\begin{center}
  \includegraphics[width=1\linewidth]{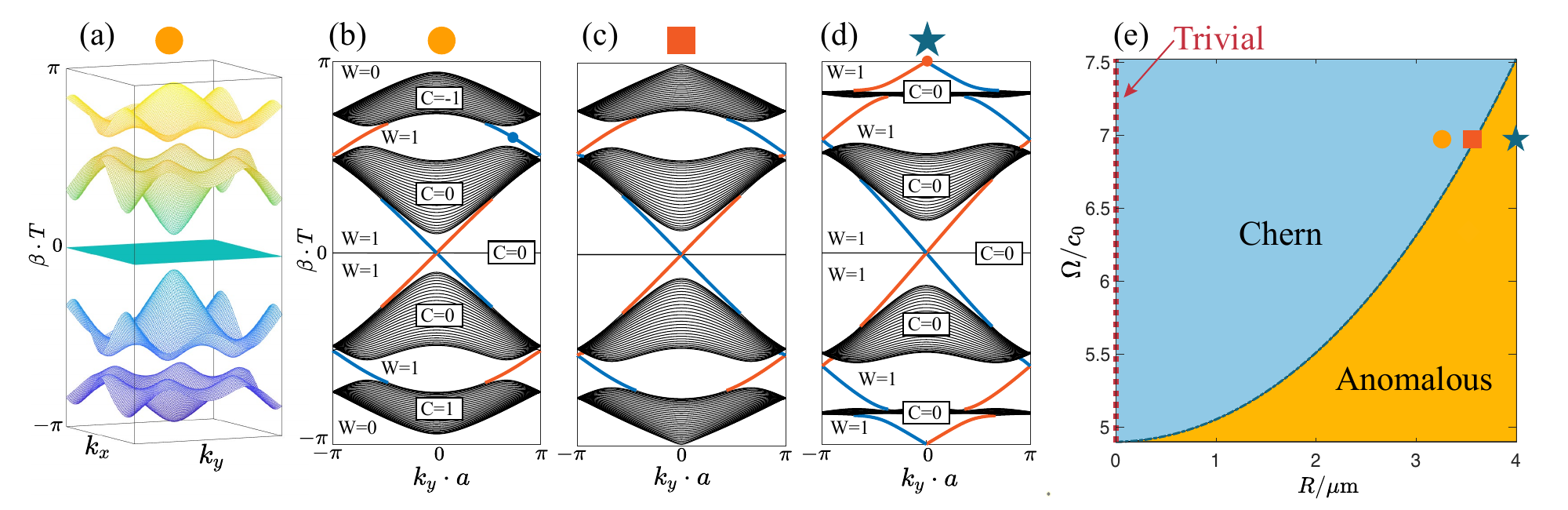}
  \caption{a) 2D Floquet-Bloch quasi-energy band structure of FCI. b, c, d) Floquet energy spectrum under open boundary condition along $x$ direction with driving parameters marked as dot, square and star in e). Chiral edge states with positive (negative) group velocity are shown in orange (blue) solid lines, while bulk states are indicated in black lines. e) Topological phase as a function of driving amplitude and frequency. The red dashed line along $R=0$ denotes trivial phase, and the blue and yellow regions denotes FCI and FATI, respectively. The parameters are chosen as $c_0=0.9\times 10^{-4}/\mathrm{\mu m}$, $d=30 \mathrm{\mu m}$ (the distance between A and B)}
  \label{f2}
\end{center}
\end{figure*}
Through introducing the evolution operator
\begin{equation}\label{eq3}
    U(\mathbf{k},z)=\tau\exp\left[-i\int^z_0dz'H(\mathbf{k},z')\right]
\end{equation}
with the time-ordering operator $\tau$, we can analyze the system with an effective Hamiltonian over one driving period, $H_{eff}(\mathbf{k})=i\mathrm{ln}[U(\mathbf{k},T)]/T$.
By diagonalizing the effective Hamiltonian, five Floquet-Bloch quasi-energy bands are obtained; see \textbf{Figure}~\ref{f2}a.
Furthermore, we can calculate the Chern number for each band, which was initially introduced to analyze quantum Hall effects.\textsuperscript{\cite{thouless1982quantized,kohmoto1985topological}}
For the parameters used in Figure~\ref{f2}a, the bottom and top bands have nonzero Chern numbers ($\pm 1$) while other bands have zero Chern numbers.
Consequently, we find that CESs appear in several gaps when the open boundary condition is set along the $x$ direction.
In detail, the number of pairs of CESs in the gap matches the sum of the Chern numbers below, which is consistent with the bulk-edge correspondence.\textsuperscript{\cite{laughlin1981quantized,halperin1982quantized,hatsugai1993chern,schulz2000simultaneous}}
Fixing the driving frequency and increasing the driving amplitude, we find that the Floquet energy gap of the bottom and top bands first closes and then reopens; see Figures~\ref{f2}c,d.
Accompanied by their Chern numbers changing from $\pm 1$ to $0$, we can identify the topological phase transition.
Even though the Chern numbers of all bands are zero, there are CESs in the energy gaps [Figure~\ref{f2}d].
This means the existence of an FATI.
To understand the topological origin of the CESs, we calculate the winding numbers in all energy gaps (see \zuomw{Supporting Information}), which all turn out to be $\mathrm{W}=1$.\textsuperscript{\cite{rudner2013anomalous}}
Then, the number of pairs of CESs in the individual gap is equal to the corresponding winding number. To understand the topological origin of the CESs, we calculate the winding numbers of the energy gaps by means of evolution operator $U(\mathbf{k},z)$ over a whole driving period (see \zuomw{Supporting Information}), in contrast to the Chern numbers obtained from the effective Hamiltonian.\textsuperscript{\cite{rudner2013anomalous}} The resulting values of $\mathrm{W}=1$ are associated with a pair of CESs in every individual gap.
In both FCI and FATI, the Chern number for a certain band is the difference between the winding numbers in its upper and lower gaps, $\mathrm{C}=\mathrm{W}_{upper}-\mathrm{W}_{below}$.

Through calculating the Chern number of the bottom band, we give the topological phase diagram in the plane of the driving amplitude $R$ and the driving frequency $\Omega$; see Figure~\ref{f2}e.
In the absence of modulation ($R=0$), the system is trivial; see the dash red line.
Increasing the driving amplitude $R$, the system becomes an FCI or FATI, depending on the driving frequency $\Omega$. 
The phase boundary between FCI and FATI is determined by the closing of Floquet energy gap between the bottom and the top bands.
The corresponding parameters for Figures~\ref{f2}b-d fall into different phases or on the boundary in the topological phase diagram. 
The energy spectra shown in these panels confirm the obtained phase boundaries and the intriguing bulk boundary correspondence of our system.

\subsection{Chiral edge states}
\begin{figure*}[!htp]
\begin{center}
  \includegraphics[width=0.5\linewidth]{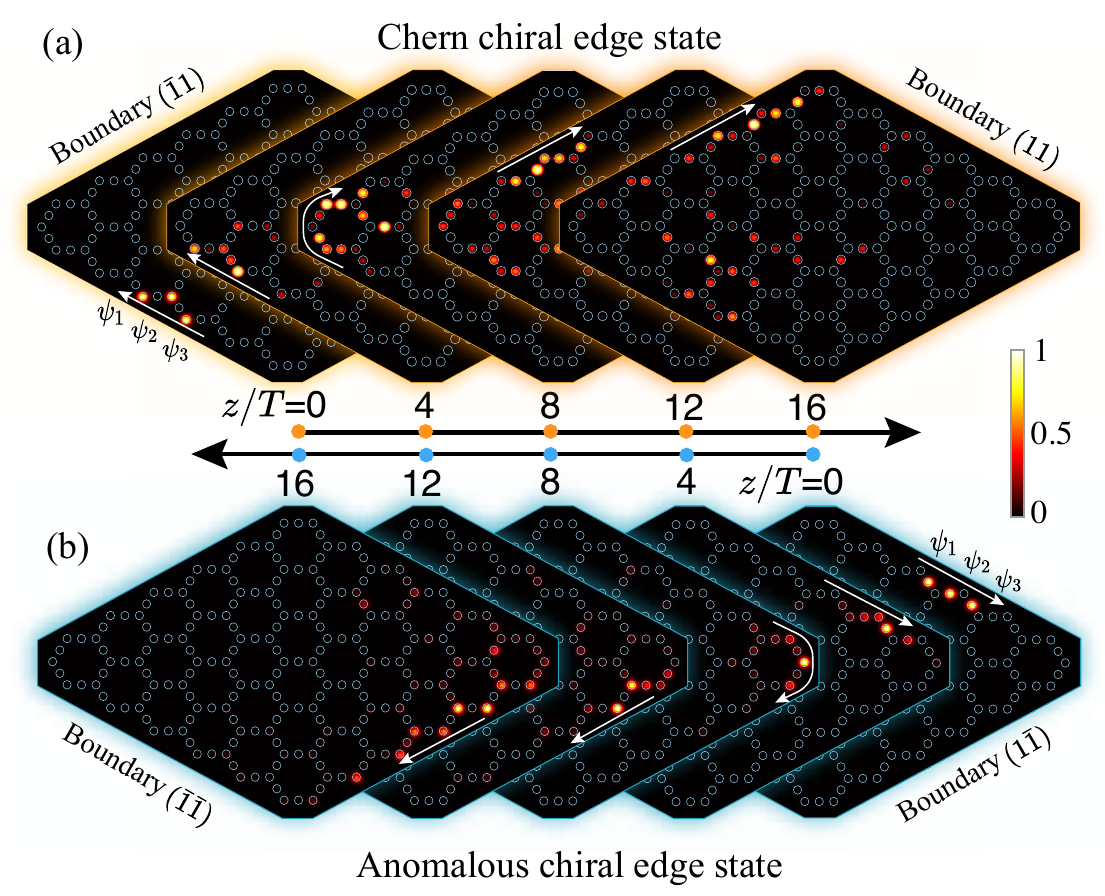}
  \caption{a, b) Density distributon of chiral edge states at the slices of $(z=0,~4T,~8T,~12T,~16T)$ in the $z$ direction in the case of FCI [marked as dot in Figure~\ref{f2}e] and  FATI [marked as star in Figure~\ref{f2}e], respectively. The white arrows guide the direction of CESs. The other parameter are chosen as $N_1=N_2=7$.}
  \label{f3}
\end{center}
\end{figure*}
In both FCI and FATI, there are CESs in all gaps with group velocity being either positive or negative; see red and blue lines in Figures~\ref{f2}b-d, respectively.
The CESs with a positive (negative) group velocity are mainly localized in the honeycomb (Kagome) sublattice along the boundaries.
We numerically simulate the transport along the boundaries shaped in rhombus with $N_1\times N_2$ cells (see \zuomw{Supporting Information}).
In the case of FCI, the initial light field $\psi(0)$ is injected into the three sites of honeycomb sublattice in the boundary $(\bar{1}\bar{1})$,
$|\psi(0)\rangle=|\psi_1\rangle e^{i\Phi}+|\psi_2\rangle+|\psi_3\rangle e^{-i\Phi}$, where $|\psi_{1,2,3}\rangle$ are the Gaussian modes marked at \textbf{Figure}~\ref{f3}a, and the phase difference is $\Phi=2\pi/3$. 
The initial state can be easier to excite and has a large overlap with the CES marked by blue dot in Figure~\ref{f2}b.
Figure~\ref{f3}a shows the light intensity distribution at different slices ($z=0, 4T, 8T, 12T, 16T$) in the $z$ direction.
We find that the light wavepacket can propagate clockwise along the boundary and smoothly navigate the corner of rhombus. 
Even after long-distance propagation ($z=16T$), CES remains localized at the boundary, demonstrating the robustness of topological transport. 
Note that a small fraction of the light is diffracted into the bulk, which is related to tiny bulk components in the initial light field. 
In the case of FATI, the initial light field $|\psi(0)\rangle$ with $\Phi=0$ is injected at the Kagome sublattice in the boundary $(11)$ to excite the CES marked by the orange dot in Figure~\ref{f2}d. 
Similarly, CES can also propagate clockwise along the boundary and pass through the corner to the other boundary; see Figure~\ref{f3}b.
The CES in long-distance propagation ($z=16T$) almost does not diffract into the bulk.
Since we cannot distinguish FCI from FATI solely using CES, below we study the bulk behaviors in response to 2D spatial gradients of refractive indices.

\subsection{2D Bloch oscillations}
\begin{figure*}[!htp]
\begin{center}
  \includegraphics[width=0.5\linewidth]{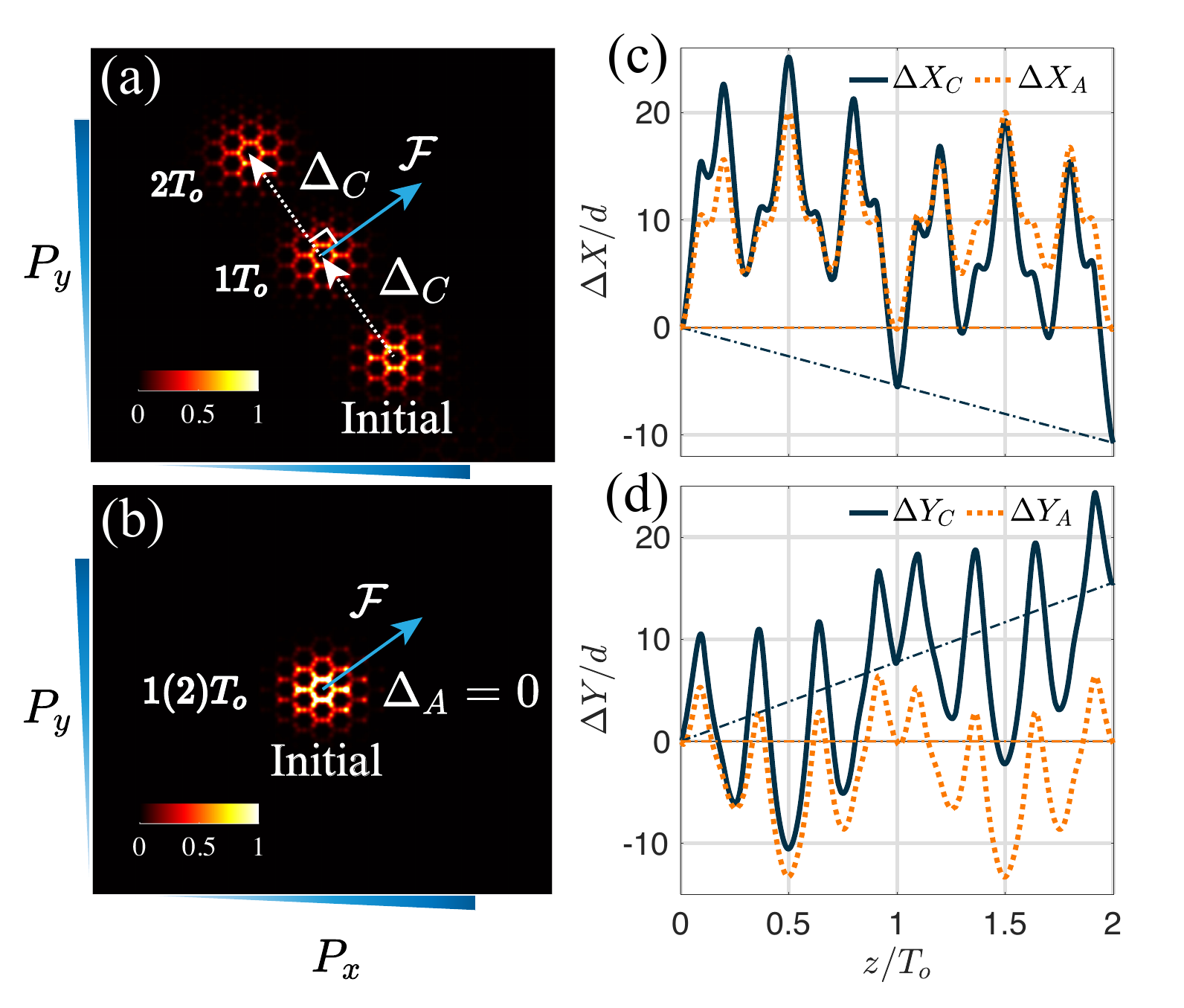}
  \caption{Bloch oscillations in FCI and FATI. a, b) Density distribution of  wavepacket of light field  at the slices $(z=0, T_o, 2T_o)$ along the $z$ direction in the case of FCI [marked as dot in Figure~\ref{f2}e] and FATI [marked as star in Figure~\ref{f2}e], respectively.
 The direction of external force $\mathcal F$ is marked by the blue arrow.
 c, d) Mean position of the wavepacket in the $x$ and $y$ directions as a function of propagation distance, respectively. The solid lines and dashed yellow lines correspond to  FCI and FATI, respectively. The other parameters are chosen as $P_x=0.52\times 10^{-5}/\mathrm{\mu m},P_y=0.21\times 10^{-5}/\mathrm{\mu m}$ and $N_1=N_2=60$. }
  \label{f4}
\end{center}
\end{figure*}
We propose to use 2D Bloch oscillations, directly related to the topological invariants, to diagnose FCI and FATI. 
We apply a 2D weak tilt potential $(P_x,P_y)$ to the large rhombus lattice, which is feasible in experiments.\textsuperscript{\cite{chang2022inhibition}}
An external force $\mathcal{F}=[\mathcal{F}_x,\mathcal{F}_y]$ is induced by the tilt potential $(P_x=\mathcal{F}_xa_x,P_y=\mathcal{F}_ya_y)$.
According to the theory of Bloch oscillations, the periods of Bloch oscillations along $x$ and $y$ directions are given by $T_o^x=2\pi/(\mathcal{F}_xa_x)$ and $T_o^y=2\pi/(\mathcal{F}_ya_y)$, respectively.
Here, $a_x=3d/2$ and $a_y=\sqrt{3}d/2$ are the lattice constants along $x$ and $y$ directions, respectively.
We impose the ratio $T_o^x/T_o^y=\eta_x/\eta_y$ with $\eta_x$ and $\eta_y$ being coprime numbers.\textsuperscript{\cite{zhu2021uncovering}}
When $T_o^x$ and $T_o^y$ are multiples of the driving period $T$, the Bloch period $T_o$ is the general multiplication of the driving period $T_o=\eta_yT_o^x=\eta_xT_o^y=nT$ (see \zuomw{Supporting Information}). 
To avoid Landau-Zener transitions between different Floquet energy bands, one may choose a sufficiently large Bloch oscillation period (such as $n=600$).
Because the potential tilt is weak, the system size has to be large enough to avoid boundary effects.
The amplitude of initial state at sublattice $j$ of cell $(m,n)$ is chosen as
\begin{equation}
    \varphi(m,n,j)=u_{1,j}(\mathbf{k})e^{i\mathbf{k}\cdot(\mathbf{r}_{mn}+\mathbf{r}_j)}e^{-{(\mathbf{r}_{mn}+\mathbf{r}_j)}^2/\sigma_0^2}, \nonumber
\end{equation}
where $u_{1,j}(\mathbf{k})$ is the periodic part of Bloch functions in the first quasi-energy band with momentum $\mathbf{k}=(0,0)$, and $\sigma_0$ is the radius of the Gaussian-like wavepacket.
The parameters are chosen as $\sigma_0=12d$, $\eta_x=2$, $\eta_y=5$, $T_x=120T$, $T_y=300T$.
The wavepacket will undergo Bloch oscillations in position space and linearly sweep the first Floquet energy band in the Brillouin zone.
If $\eta_x/\eta_y$ is large or small enough, the wavepacket can pick up Berry curvaturs along the trajectory in momentum space and effectively uniformly sweep the whole Brillouin zone.
As a consequence, the wavepacket in the position space will experience a unidirectional drift due to the anomalous group velocity associated with the Berry curvatures.
The displacement $(\Delta X_C,\Delta Y_C)$ in an overall Bloch period is related to quasi-quantized number $C_x$ and $C_y$, 
\begin{equation}
    \begin{aligned}
        \mathrm{C}_x=-\frac{\Delta X_C(T_o)}{2a_x\eta_x},\mathrm{C}_y=\frac{\Delta Y_C(T_o)}{2a_y\eta_y},
    \end{aligned}
\end{equation}
where $C_x\approx C_y\approx \mathrm{C}$ are close to the Chern number $\mathrm{C}$ of the first band.
The overall displacement given by $|\Delta_C|=\sqrt{\Delta X_C^2+\Delta Y_C^2}$ will be perpendicular to the direction of external force $\Delta_C \perp\mathcal{F}$.

In the case of an FCI, \textbf{Figure}~\ref{f4}a shows the density distribution of wavepacket on slides of $0, T_o, 2T_o$, and black solid lines in Figures~\ref{f4}c,d show the displacements along the $x$ and $y$ directions. 
The wavepacket will undergo 2D Bloch oscillations with near quantized drifts with $C_x\approx 0.92$ and $C_y\approx 0.9$, which is close to the Chern number $\mathrm{C}=1$.
In each overall cycle, the wavepacket has the same displacement $\Delta_C$ in the direction perpendicular to the applied external force. 
In contrast, in the case of FATI, the wavepacket returns to the same position as the initial state at the first and second overall Bloch periods; see Figure~\ref{f4}b.
The wavepacket undergoes 2D Bloch oscillations without displacement $(\Delta X_A=\Delta Y_A=0)$; see the dashed orange lines in Figures~\ref{f4}c,d.
This is consistent with the zero Chern number of the first Floquet energy band in an FATI. 
These results indicate that Bloch oscillations provide an efficient method for diagnosing FCI and FATI.

\section{Conclusion}
We proposed a Floquet tight-binding model that supports topological phase transitions between trivial insulator, FCI, and FATI by adjusting the driving parameters. 
As the common topological transport of CESs cannot distinguish FCI and FATI, we propose using 2D Bloch oscillations to diagnose the different Floquet topological phases. 
In stark contrast to quantum Hall effects in response to 1D Bloch oscillations, which require uniform band occupation,\textsuperscript{\cite{ke2016topological,hu2019dispersion}} our scheme does not need uniform band occupation. 
Furthermore, our scheme can be generalized to the diagnosis of other  Floquet topological phases and can be experimentally realized in other systems such as ultracold atoms and superconducting circuits.\textsuperscript{\cite{wintersperger2020realization,braun2024real,zhang2022digital}}

It is interesting to generalize our studies from linear to nonlinear systems by considering Kerr effects of light.
The study of nonlinear Floquet topological insulators\textsuperscript{\cite{smirnova2020nonlinear}} is still unexplored rich and promising, though some remarkable studies have exposed nonlinear Thouless pumps,\textsuperscript{\cite{jurgensen2021quantized,fu2022nonlinear,jurgensen2023quantized}} nonlinearity-induced topological phase transitions,\textsuperscript{\cite{leykam2016edge,maczewsky2020nonlinearity,xia2021nonlinear}} and Floquet solitons.\textsuperscript{\cite{lumer2013self,mukherjee2020observation,mukherjee2021observation,zhong2023pi}}
Although double-period Floquet solitons have been observed in the Floquet honeycomb model,\textsuperscript{\cite{mukherjee2023period}} it is an open question whether multiple-period Floquet solitons exist in the nonlinear region of our proposed systems.
Furthermore, it is worth investigating the interplay between nonlinearity, space-time symmetries, and topology in nonlinear Floquet topological insulators.

\medskip
\textbf{Supporting Information} \par 
Supporting Information is available from the Wiley Online Library or from the author.

\medskip
\textbf{Acknowledgements} \par 
This work was supported by the National Key Research and Development Program of China (Grant No. 2022YFA1404104), the National Natural Science Foundation of China (Grants Nos. 12025509, 12275365, and 92476201), the Guangdong Provincial Quantum Science Strategic Initiative (GDZX2305006, GDZX2304007, GDZX2405002), and the Natural Science Foundation of Guangdong(Grant No. 2023A1515012099).
\newpage

\begin{flushleft}
    \huge {Supporting Information for Diagnosing Floquet Chern and Anomalous Topological Insulators based on Bloch Oscillations}\\[1ex]
\end{flushleft}
\vspace{1em}

\setcounter{section}{0}
\setcounter{equation}{0}
\setcounter{figure}{0}
\setcounter{table}{0}

\renewcommand{\thesection}{S\arabic{section}}
\renewcommand{\theequation}{S\arabic{equation}}
\renewcommand{\thefigure}{S\arabic{figure}}
\renewcommand{\thetable}{S\arabic{table}}

\section{Coupling and Detuning between waveguides}
The propagation of light in photonic waveguide arrays is usually described by the paraxial Helmholtz equation,
\begin{equation}\label{eqs1}
     i\partial_{z}\psi(z)=H_0\psi(z),
\end{equation} 
where the Hamiltonian is given by
\begin{equation}
    H_0(z)=-\frac{1}{2k_{0}}\nabla^{2}-\frac{k_{0}\Delta n(z)}{n_{0}}.
\end{equation}
Here, the wavefunction $\psi(z)$ corresponds to the envelope of the electric field of light, $\mathbf{E}(z)=\psi(z)\exp(ik_0z-i\omega t)\mathbf{E}_0$, where $\mathbf{E}_0$ is a unit vector, $\omega$ is the frequency of light, and $t$ is time. 
$\nabla^{2}=\partial_{x}^{2}+\partial_{y}^{2}$ is the transverse Laplacian,  and $k_0=2\pi n_0/\lambda$ is the wave number in the background medium with $n_0$ being the background refractive index and $\lambda$ being the wavelength of the injected field. 
Without loss of generality, in this work the wavelength $\lambda$ is chosen as $633 \mathrm{nm}$, and the background refractive index $n_0$ is chosen as $1.45$. 
We can design the relative refractive index as the sum of potentials of straight honeycomb sublattice ($V_h$) and curved Kagome sublattice [$V_k(z)$],
\begin{equation}
    \Delta n(z)=V_h+V_k(z).
\end{equation}
The potentials for individual waveguides are given by $V_0\exp(-|\mathbf{r}(z)|^2/\sigma^2)$, where the effective waveguide radius is $\sigma=3.5\mathrm{\mu m}$, $\mathbf{r}=(x,y)$, and $V_0=1.2\times 10^{-3}$, respectively.  
This can be easily achieved with mature laser-direct-write techniques.\textsuperscript{\cite{szameit2010discrete}}

\begin{figure*}[!htp]
\begin{center}
 \includegraphics[width=0.9\linewidth]
 {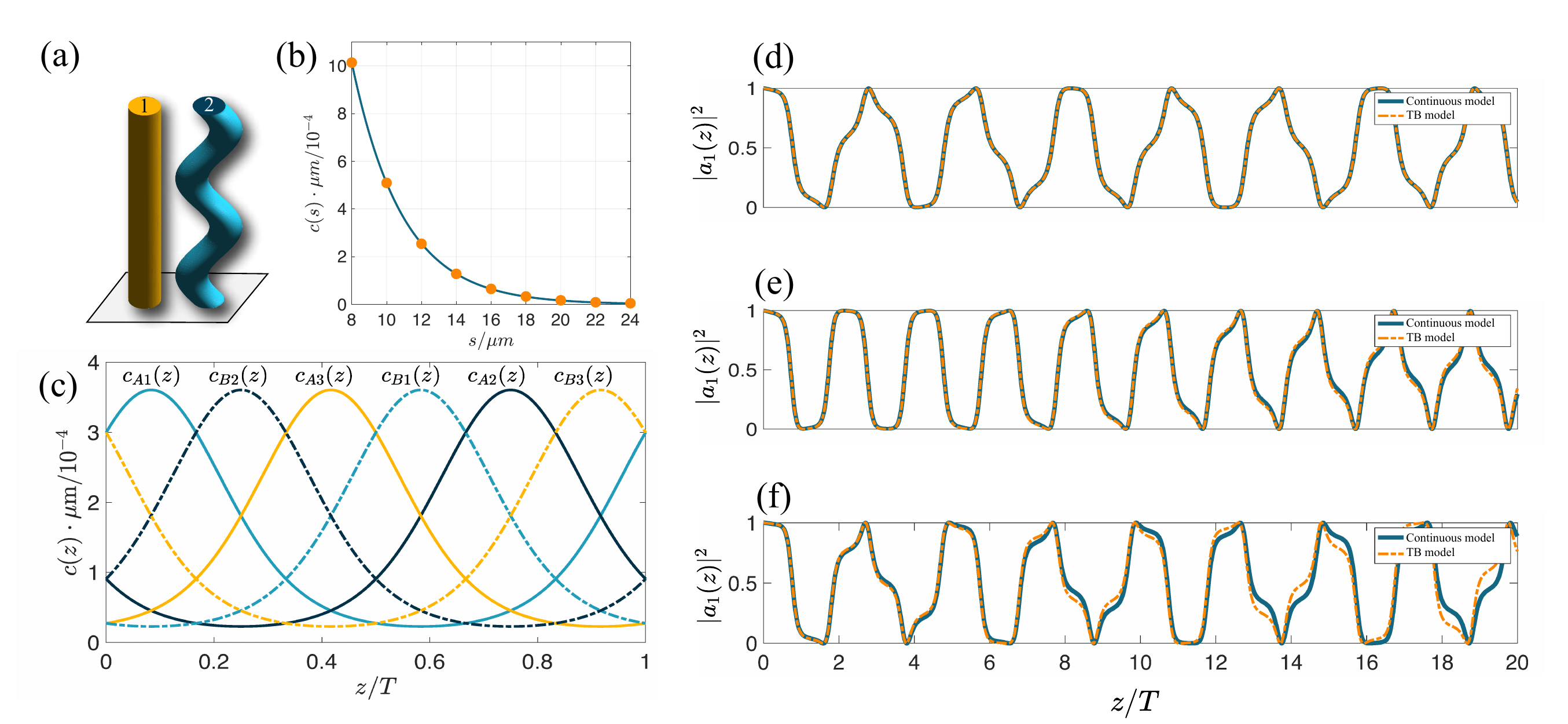}
    \caption{a) The two-mode continuous model involving a straight and a curved waveguide. b) The curve of the coupling strength $c(s)$ is shown, where the yellow dots represent the numerical calculation results, and the blue solid line denotes the fitted curve. c) The coupling curves between different straight and curved waveguides, corresponding to the blue star in Figure 2e of the main text. d-f) The probabilities at the first waveguide as a function of propagation distance in the Chern insulator ($R=3.5\mathrm{\mu m},\Omega=2\pi/\mathrm{cm}$), anomalous topological insulator ($R=4\mathrm{\mu m},\Omega=2\pi/\mathrm{cm}$) and lattice under stronger and faster modulation ($R=4\mathrm{\mu m},\Omega=2.174\pi/\mathrm{cm}$, respectively. The solid blue and dashed yellow lines denote the results obtained via continuous model and tight-binding (TB) model, respectively.  }
 \label{fs1}
 \end{center}
\end{figure*}

Because the coupling strength between waveguides exponentially decays with waveguide spacing, we only keep the nearest-neighboring coupling between waveguides and ignore the longer-range couplings.
The basic coupling and detuning of the lattices can be extracted by analyzing the nearest-neighboring curved waveguide and the straight waveguide; as depicted in \textbf{Figure}~\ref{fs1}a. 
Here, we derive parameters of a tight-binding model from a two-mode continuous model based on straight and curved waveguides. 
The light field transport in the curved waveguide follows a longer optical path compared to the straight waveguide, causing a phase difference between the wave functions of the two fundamental Gaussian modes.
We express the wave function of the two-mode system as,
\begin{equation}\label{eqs2}
    |\psi\rangle=b_1(z)|\psi_1\rangle+b_2(z)e^{ik_0\Delta z}|\psi_2\rangle.
\end{equation}
Here, $\Delta z=\Delta z_1+\Delta z_2$ represents the optical path difference between the two waveguides along the $z$ direction, with $\Delta z_1=R^2\Omega^2z/4$ and $\Delta z_2=R^2\Omega\sin(2\Omega z)/8$. %
$|\psi_1\rangle$ and $|\psi_2\rangle$ are orthogonal fundamental Gaussian modes localized in the first and second waveguides, $\langle\psi_i|\psi_j\rangle=\delta_{ij}$.
They can be constructed by two lowest eigenstates of $H_0$ via 
\begin{eqnarray}
     |\psi_1\rangle&=&(|\phi_g\rangle+|\phi_e\rangle)/\sqrt{2},\\
     |\psi_2\rangle&=&(|\phi_g\rangle-|\phi_e\rangle)/\sqrt{2},
\end{eqnarray}
where $|\phi_g\rangle$  and $|\phi_e\rangle$ are the ground state and excited state, $H_0|\phi_s\rangle=\varepsilon_s|\phi_s\rangle$ ($s=g,e$), respectively. They can be obtained through numerical calculation of virtual time evolution.
From \textbf{Equations}~\eqref{eqs1} and \eqref{eqs2}, disregarding extremely small terms, we obtain the two-mode tight-binding model,
\begin{align*}
    i\partial_zb_1(z)&=b_1(z)\varepsilon_0+b_2(z)c(z)e^{ik_0\Delta z},\\
    i\partial_zb_2(z)&=b_1(z)c(z)e^{ik_0\Delta z}+b_2(z)[\varepsilon_0+k_0\partial_z\Delta z],
\end{align*}
with $\varepsilon_0$ being the constant on-site energy. Hence, the tight-binding Hamiltonian is given by,
\begin{align*}
    H(z)=\varepsilon_0\Psi_1^\dagger\Psi_1+(\varepsilon_0+k_0\partial_z\Delta z)\Psi_2^\dagger\Psi_2-c(z)e^{ik_0\Delta z}\Psi_1^\dagger\Psi_2+\mathrm{h.c.},
\end{align*}
where $\Psi_j^{\dagger}$ and $\Psi_j$ are the creation and annihilation operators of a Gaussian mode at the $j$th site. The maximum of $\Delta z_2$ given by $k_0R^2\Omega/8$ is negligibly small and can be disregarded, leading to $\Delta z\approx z_1$. 
In this context, the $\Psi_1$ and $\Psi_2$ can be gauge transformed into $\widetilde\Psi_1=\Psi_1$ and $\widetilde\Psi_2=\Psi_2e^{ik_0\Delta z_1}$, and the Hamiltonian can be written as
\begin{equation} \label{HamTwo}
    H(z)=\varepsilon_0\widetilde\Psi_1^\dagger\widetilde\Psi_1+[\varepsilon_0+\Delta\varepsilon(z)]\widetilde\Psi_2^\dagger\widetilde\Psi_2-c(z)\widetilde\Psi_1^\dagger\widetilde\Psi_2+\mathrm{h.c.},
\end{equation}
where the detuning $\Delta\varepsilon(z)=k_0R^2\Omega^2(\cos^2(\Omega z)-1/2)/2$, and the onsite-energy $\varepsilon_0$ and coupling $c(z)$ can be numerically calculated by
\begin{align*}
   & c(z)=\langle \psi_{1(2)}|H_0(z)|\psi_{2(1)}\rangle=\frac{\varepsilon_e(z)-\varepsilon_g(z)}{2},\\
    &\varepsilon_0=\langle\psi_{1(2)}|H_0(z)|\psi_{1(2)}\rangle=\frac{\varepsilon_g(z)+\varepsilon_e(z)}{2}.
\end{align*}
We obtain the coupling strength $c(z)$ by varying the distance between the two waveguides and fit it as a function of the waveguide spacing. The envelope functions of both $|\psi_1\rangle$ and $|\psi_2\rangle$ are exponential, and the coupling strength decays exponentially as the waveguide spacing increases. 
The fitted function for the coupling strength $c(z)\cdot\mathrm{\mu m}=0.016\exp[-0.345s(z)/\mathrm{\mu m}]$ where $s(z)$ indicates the distance between two waveguides; see Figure~\ref{fs1}b.

To validate the accuracy of parameters for the tight-binding model, we compare the transport dynamics governed by the tight-binding model and its original continuous model.\textsuperscript{\cite{PhysRevLett.110.243902}}
The amplitude of detuning $\Delta\varepsilon(z)$ can be further ignored in the tight-binding model, which is much smaller compared to the coupling strength. 
The on-site energy $\varepsilon_0$ is a constant value which only contributes to the overall dynamical phase, and we set it  as $\varepsilon_0=0$ for convenience. 
We numerically solve the following Schr\"odinger equations,
\begin{align*}
    i\partial_z|b(z)\rangle=H(z)|b(z)\rangle,
\end{align*}
and
\begin{align*}
    i\partial_z |\psi(z)\rangle= H_0(z)|\psi(z)\rangle.
\end{align*}
The initial states are set as Gaussian wavefunction located at the first waveguide in real space for the first case and $|\psi(0)\rangle=[1,0]^{T}$ for the second case, respectively.
We compare the probability density of the light field at the first waveguide along the propagation direction, $|b_1(z)|^2$ and $|\langle \psi_1|\psi(z)\rangle|^2$. 
We have selected three sets of parameters $(R,~\Omega)=(3.5\mathrm{\mu m},~\Omega=2\pi/\mathrm{cm})$, $(4\mathrm{\mu m},~2\pi/\mathrm{cm})$ and $(4\mathrm{\mu m},~2.174\pi/\mathrm{cm})$. The first two sets of parameters correspond to the Chern topological phase and anomalous topological phase within the topological phase diagram in the main text, respectively. 
However, the last one corresponds to the parameters beyond the scope of the topological phase diagram in the main text.
Within twenty driving periods, the probability densities calculated by the tight-binding model and the original continuous models are highly consistent with each other [Figures~\ref{fs1}d,e], confirming the reliability of our calculations. 
However, it is important to note that the tight-binding approximation has limitations.
When the driving amplitude and frequency are too large, the results obtained by the two models become mismatched after long-distance transport; see Figure~\ref{fs1}f. 
To avoid possible errors caused by the breakdown of tight-binding approximations, we only consider parameters within the range shown in Figure~2 of the main text.

\section{Tight-binding Hamiltonian and and topological invariants}\label{sup2}
\subsection{Hamiltonian}
To simulate light transport, we can extend the two-lattice model~\eqref{HamTwo} to honeycomb-Kagome lattice in real space under open boundary condition,
\begin{equation}\label{eqs4}
    \begin{aligned}
       H(z)=&-\sum_{m=1}^{N_1}\sum_{n=1}^{N_2} \big[ c_{B,2}(z)\hat\psi_{B,m,n}^{\dagger}\hat\psi_{2,m,n}+\sum_{j=1,2,3}c_{A,j}(z)\hat\psi_{A,m,n}^{\dagger}\hat\psi_{j,m,n}+\mathrm{h.c.}\big] 
       \\&-\sum_{m=1}^{N_1-1}\sum_{n=1}^{N_2-1}\left[c_{B,1}(z)\hat\psi_{B,m,n}^{\dagger}\hat\psi_{1,m+1,n}+c_{B,3}(z)\hat\psi_{B,m,n}^{\dagger}\hat\psi_{3,m,n+1}+\mathrm{h.c.}\right].
\end{aligned}
\end{equation}
Here, the onsite energies are so weak that they can be safely neglected. 
$\hat\psi_{s,m,n}^\dagger$ and $\hat\psi_{s,m,n}$ are the creation and annihilation operators of Gaussian modes in the $s$ type of waveguide in the $(m,n)$ cell. 
$N_1$ and $N_2$ are the numbers of cells in two different directions.
$c_{s,s'}$ are the coupling strengths between the $s$ and $s'$ types of waveguides.
Under periodic boundary conditions, the wave function of the light field in the lattice can be expressed as a Bloch wave,
\begin{equation}
    \psi(\mathbf{k})=\frac{1}{\sqrt{N_1\cdot N_2}}\sum_{m=1}^{N_1}\sum_{n=1}^{N_2}\sum_{s=1,2,3,A,B}\psi_{s,m,n}e^{i\mathbf{k}\cdot \mathbf{R}_{m,n}^{s}},
\end{equation}
where $R_{m,n}^{s}$ is the vector of the $s$ type lattice site. 
To derive the Bloch waves,  we need to perform a Fourier transform of \textbf{Equation}~\eqref{eqs4}, by introducing the creation and annihilation operators ($\psi_{s,\mathbf{k}}^{\dagger}$, $\psi_{s,\mathbf{k}}$) in the momentum space which is related to those in the real space
\begin{eqnarray}
    \hat\psi_{\Lambda(j),m,n}^{\dagger}&=&\frac{1}{\sqrt{N_1\cdot N_2}} \sum _{\mathbf{k}}\psi_{\Lambda(j),\mathbf{k}}^{\dagger}e^{i\mathbf{k}\cdot\mathbf{R}^{\Lambda(j)}_{m.n}}, \\ \nonumber
    \hat\psi_{\Lambda(j),m,n}&=&\frac{1}{\sqrt{N_1\cdot N_2}}\sum_{\mathbf{k}}\psi_{\Lambda(j),\mathbf{k}} e^{-i\mathbf{k}\cdot\mathbf{R}^{\Lambda(j)}_{m.n}}.
\end{eqnarray} 
In this way, we get a more concise Hamiltonian in the momentum space
\begin{equation}
H(\mathbf{k},z) = - \sum_{\Lambda=A,B}\sum_{j=1,2,3}c_{\Lambda,j}(z)\psi_{\Lambda,\mathbf{k}}^\dagger\psi_{j,\mathbf{k}}\mathrm{e}^{i\mathbf{k}\cdot\boldsymbol{\delta}_{\Lambda,j}}+
 \mathrm{h.c.}.
\end{equation}

\subsection{Topological invarants: Chern number and Winding number}
The two-dimensional (2D) quasi-energy band such as Figure2(a) in the main text can be obtained by $H_{eff}(\mathbf{k})|u_{n,\mathbf{k}}\rangle=\beta_{n,\mathbf{k}}|u_{n,\mathbf{k}}\rangle$. 
With the Floquet Bloch states $u_{n,\mathbf{k}}\rangle$, the Chern number of $n$-th band can be calculated by 
\begin{equation}
    C_n=\frac{1}{2\pi}\int\int_{FBZ}dk_xdk_y \cdot\nu(k_x,k_y),
\end{equation}
where $\nu(k_x,k_y)=\mathrm{Im}(\langle\partial_{k_y}u_n|\partial_{k_x}u_n\rangle-\langle\partial_{k_x}u_n|\partial_{k_y}u_n\rangle)$ is the Berry curvature.\textsuperscript{\cite{ke2016topological}}

In 2D anomalous Floquet topological insulators, all band Chern numbers are $0$ and cannot be used as topological invariants. 
In such systems, the gap winding number\textsuperscript{\cite{rudner2013anomalous}} is usually calculated by
\begin{equation}
\begin{aligned}
        \mathrm{W}[U]=\frac{1}{8\pi^{2}}\int\mathrm{d}z\mathrm{d}k_{x}\mathrm{d}k_{y}\cdot \mathrm{Tr}\left(U^{-1}\partial_{z}U\big[U^{-1}\partial_{k_{x}}U,U^{-1}\partial_{k_{y}}U\big]\right).
\end{aligned}
\end{equation}

The number of chiral edge states(CESs) in the $\beta$ gap is equal to the winding number $N_{\mathrm{edge}}=\mathrm{W}[U_\beta]$ with $U_\beta$ defined as
\begin{equation*}
    \begin{aligned}
    \left.U_\beta(\mathbf{k},z)=\left\{\begin{array}{cc}U(\mathbf{k},2z)&0\le z\le\frac{T}{2}\\V_\beta(\mathbf{k},2T-2z)&\frac{T}{2}\le z\le T.\end{array}\right.\right.
\end{aligned}
\end{equation*}
Here, $V_{\beta}(\mathbf{k},z)=\exp(-iH_{eff}(\mathbf{k})z)$ and the branch cut of  logarithm of the effective Hamiltonian is given by
\begin{align*}
    \begin{aligned}&\log e^{-i\beta T+i0^-}=-i\beta T,\\&\log e^{-i\beta T+i0^+}=-i\beta T-2\pi i.\end{aligned}
\end{align*}

\section{Topological Chiral edge states} \label{sup3}
In this section, we calculate the topological CESs under two types of boundary conditions.
The first case takes open boundary condition along $x$ direction and periodic boundary condition along $y$ direction, while the second case takes open boundary condition along both the $x$ and $y$ directions.
The distributions of topological CESs are presented for both cases. 
In the first case, the lattice turns into a ribbon lattice, and we can observe the distribution of CESs with certain quasi-momentum $k_y$ and group velocities across different gaps. 
In the second case, we can observe that CESs with distinct group velocities occupy different sublattices.
This also facilitates selecting initial positions to excite certain CESs for studying their light propagation.

\subsection{Topological Chiral edge states in a one-dimensional ribbon lattice}\label{sup3.1}
\begin{figure}[htbp]
\centering
 \includegraphics[width=1\linewidth]
 {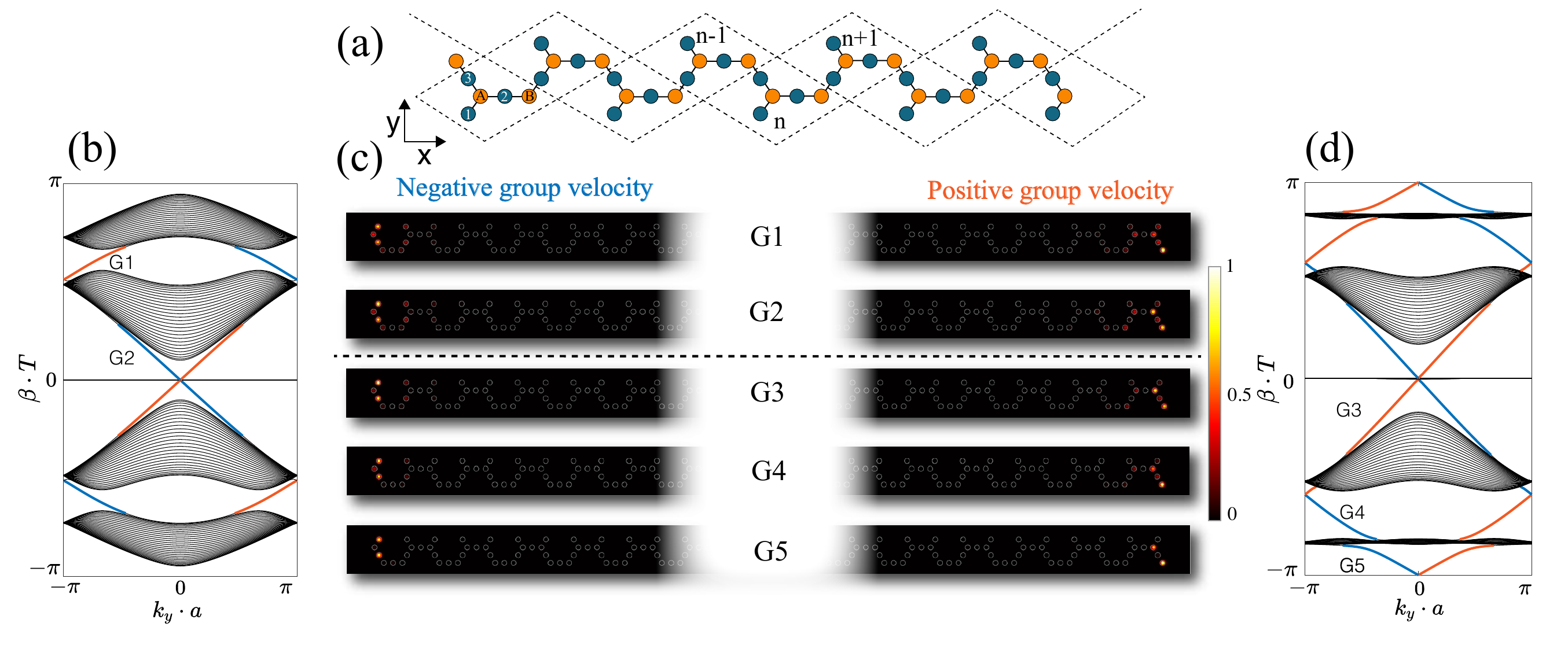}
 \caption{Chiral edge states (CESs) in different topological phases. a) Schematics of a one-dimensional ribbon chain. 
 b) The Floquet energy band of the Chern topological insulator (marked as dot in \textbf{Figure} 2e of the main text). c) CESs with different group velocities, energy gaps and topological phases.
 The upper two rows show the CESs in the first and second gaps of the Chern topological insulators, marked by $G1$ and $G2$ in b), respectively.
The lower three rows show the CESs in the third, fourth and fifth gaps in the anomalous topological phase, marked by $G3$, $G4$ and $G5$ in d), respectively.
From the top to the bottom, the CESs with positive and negative group velocities have quasi-momentum $k_y\cdot a= \pi[\pm\frac{2}{3},\pm\frac{1}{5},\mp\frac{1}{5},\mp\frac{2}{3},0]$, respectively.
d) The band of the anomalous topological insulator (marked as star in Figure 2e of the main text). }
\label{fs2}
\end{figure}

By imposing open boundary condition along the $x$ direction and periodic boundary condition along the $y$ direction, we can make partial Fourier transformation of the Hamiltonian~\eqref{eqs4} along the $y$ direction and derive a ribbon model depending on quasi-momentum $k_y$ and propagation direction $z$,
\begin{equation} \label{Hamkz}
    H(k_y,z)=-\sum_{n=1}^N\left(\sum_{j=1,2,3}c_{A,j}(z)\hat\psi_{A,n}^\dagger\hat\psi_{j,n}e^{ik_y\zeta_{A,j}}+c_{B,2}\hat\psi_{B,n}^\dagger\hat\psi_{2,n}e^{ik_y\zeta_{B,2}}\right)
    -\sum_{n=1}^{N-1}\sum_{j=1,3}c_{B,j}(z)\hat\psi_{B,n}^\dagger\hat\psi_{j,n+1}e^{ik_y\zeta_{B,j}}+\mathrm{h.c.},
\end{equation}
As depicted in \textbf{Figure}~\ref{fs2}a, each cell contains $5$ lattices, with the left and right edges ending at the $B$ and $A$ sites. 
Here, $N$ is the cell number of the chain and  $\zeta_{\Lambda,j}$ is the vector in the $y$ direction between $\Lambda$ and $j$ sites. The vectors are given by $\zeta_{A,1}=\zeta_{B,3}=-a/4$, $\zeta_{A,2}=\zeta_{B,2}=0$, and $\zeta_{A,3}=\zeta_{B,1}=a/4$. 
By diagonalizing the Hamiltonian~\eqref{Hamkz}, we can obtain quasi-energy spectra as functions of $k_y$ in Figures~\ref{fs2}b, d and Figures~2b,d of the main text.

This model can demonstrate different topological phases by varying driving amplitudes and frequencies. 
Figures~\ref{fs2}b, d display the Floquet energy bands corresponding to the Chern topological phase and the anomalous topological phase, respectively. 
Unlike the model in Reference~\cite{pyrialakos2022bimorphic}, in our proposed lattice CESs with the same group velocity in different gaps and different topological phases are localized at the same type of sublattices.
As shown in Figure~\ref{fs2}c, the upper two rows correspond to the energy bands of the Chern topological phase (Gaps $\mathrm{G1}$ and $\mathrm{G2}$), while the lower three rows correspond to the gaps $\mathrm{G3-G5}$ of the anomalous topological phase.
Because the spectra are symmetric about the zero energy,  we only show CESs in the upper two gaps of the Chern topological phase and the lower three gaps of the anomalous topological phase.
Interestingly, CESs with a positive group velocity are mostly localized in the Kagome lattices around the left boundary, while CESs with a negative group velocity are mostly localized at the honeycomb lattices around the right boundary. 
%
%

\subsection{Topological Chiral edge states under open boundary conditions} \label{sup3.2}
\begin{figure*}[!htp] 
\begin{center}
 \includegraphics[width=1\linewidth]
 {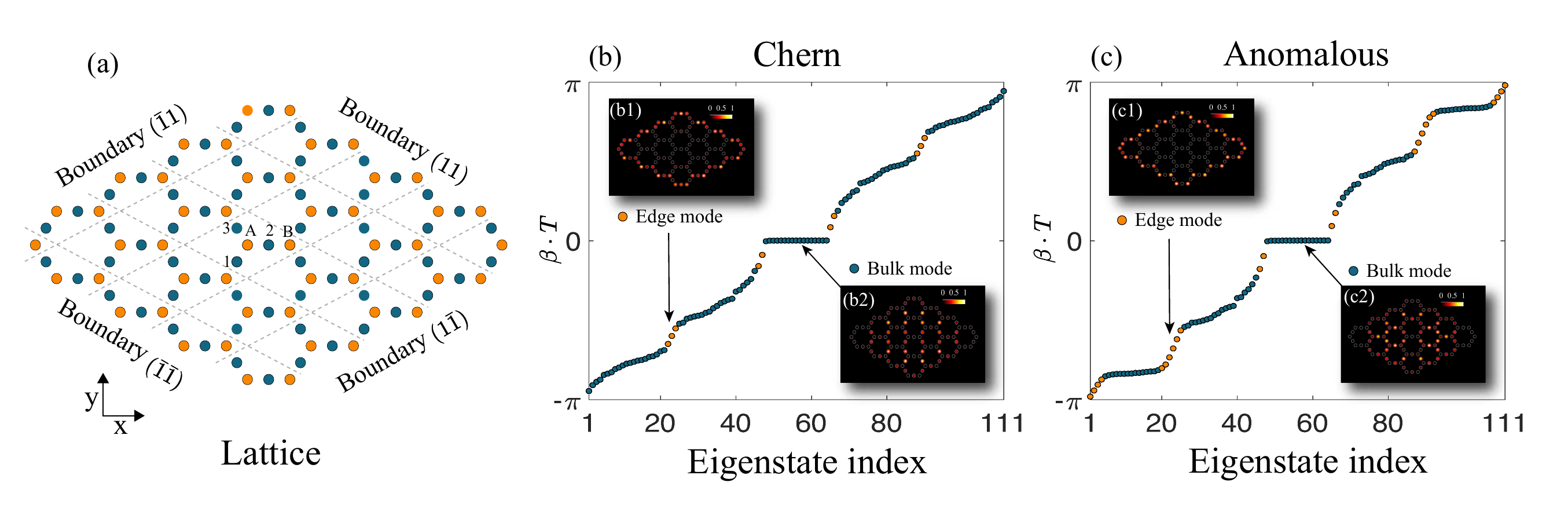}
 \caption{a) honeycomb-Kagome lattice under open boundary conditions. b) and c) The spectrum of Chern topological phase ($R=3.5\mathrm{\mu m},\Omega=2\pi/\mathrm{cm}$) and anomalous topological phase ($R=4\mathrm{\mu m},\Omega=2\pi/\mathrm{cm}$), respectively.  The insets b1) and b2) show the distribution of chiral edge states and bulk states with energy connected the the point in the spectrum of Chern topological insulator. The insets c1) and c2) are similar to b1) and b2), but in anomalous topological insulator. }
\label{fs3}
\end{center}
\end{figure*}

We present some typical states in a rhombic lattice with $111$ sites under open boundary conditions along both $x$ and $y$ directions; see \textbf{Figure}~\ref{fs3}a.
The quasi-energy spectrum and Floquet states can be obtained by diagonalizing the effective Hamiltonian(\ref{eqs4}); see Figure~\ref{fs3}b and~\ref{fs3}c for the Chern insulator and the anomalous topological insulator, respectively.
In contrast to the Chern insulator, the anomalous topological phase exhibits CESs in the $\pi$ gap. 
Figures~\ref{fs3}b1,c1 and~\ref{fs3}b2,c2 show the spatial distribution of CESs and the bulk states in the two topological phases, respectively. 
These Floquet states have quasi-energies that are connected to points of the spectrum by arrows.
We observe that the density of the CESs is mainly localized in the Kagome sublattices  along the boundaries $(\bar 1 1)$ and $(11)$ and in the honeycomb sublattices along the boundaries $(\bar 1\bar 1)$ and $(1\bar1)$.
The degree of localization of CESs is more obvious in anomalous topological phases. 
Furthermore, the density distributions of both the CESs and bulk states are mirror-symmetrical about the $y$-axis. 
This indicates that boundaries $(\bar 1 1)$ and $(11)$ are equivalent based on the miror symmetry, as are boundaries $(\bar 1\bar 1)$ and $(1\bar1)$. 
When performing numerical simulations of the transport of chiral edge states, we can refer to the distribution of the CESs to select and excite the initial states at different positions and different boundaries, accordingly.

\section{Quasi-quantized  displacement in 2D tilted honeycomb-Kagome lattice}\label{sup4}

 \begin{figure*}[!hbp]
\begin{center}
 \includegraphics[width=0.9\linewidth]
 {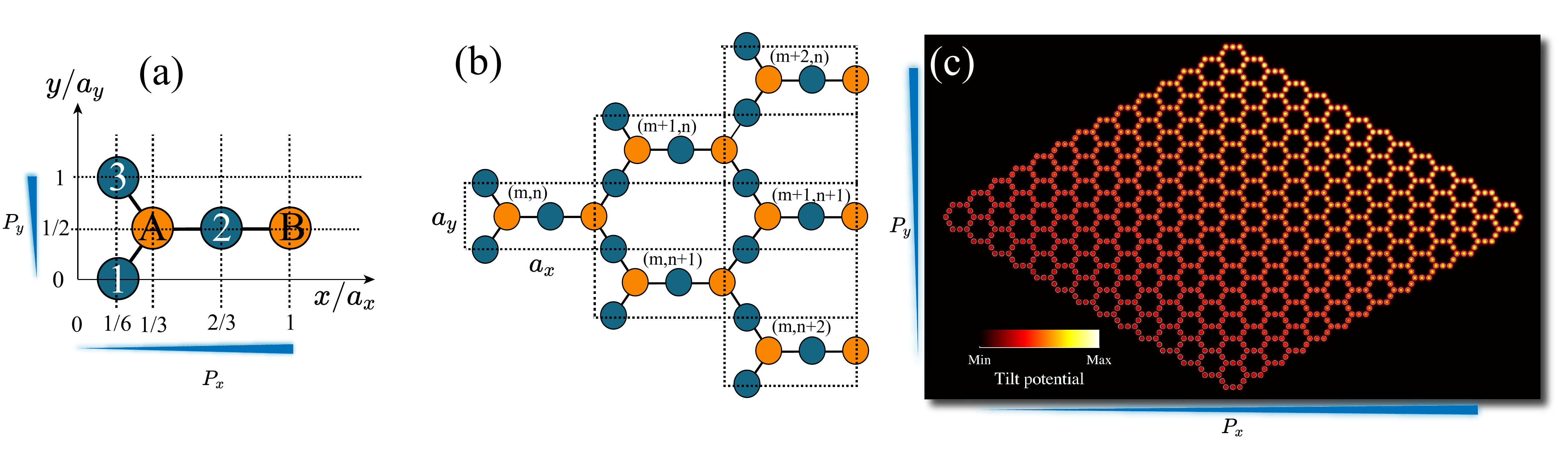}
 \caption{ Schematics of the two-dimensional tilted potential. a) The coordinates of the lattice sites within a unit cell. 
 b) The cell indices in the zoomed-in structures. 
 c) The change of tilted potential in the lattice.}
\label{fs4}
\end{center}
\end{figure*}

Here, we show how to use 2D Bloch oscillations to realize the quasi-quantized drift in a 2D tilted honeycomb-Kagome lattice.
The tilted potential is realized by introducing a gradient of refractive indices along both the $x$ and $y$ directions.
The tilted potential makes the wavepacket uniformly sweep the Brillouin zone and accumulate the Berry curvatures, which result in nearly quantized drift of Bloch oscillations in both $x$ and $y$ directions. 
We set a weak tilted potential, in order to be as close to adiabatic sweeping of the Floquet energy band as possible.
However, weak tilted potential also means large expansion of wavepacket in  Bloch oscillations.
To avoid boundary effects, we also need to set a large system size.

Before proceeding to analyze the displacement in the Bloch oscillations, we will first show how to construct the tilted honeycomb-Kagome lattice.
The lattice contains 3600 unit cells, and the position of the $\{m,n\}$th cell is given by the vector, $\mathbf{R}_{m,n}=m\mathbf{R}_1+n\mathbf{R}_2$ with basis $\mathbf{R}_1=[a_x,a_y]$ and $\mathbf{R}_2=[a_x,-a_y]$.
We introduce a 2D weak tilting potential,
\begin{equation}
\begin{aligned}
    H_p&=\sum_{m=1}^{N_1}\sum_{n=1}^{N_2}\sum_{j=A,B,1,2,3}(M_x+q^x_j,M_y+q^y_j)\cdot \mathbf{P}\psi^{\dagger}_{j,m,n}\psi_{j,m,n} ,
    \end{aligned}
\end{equation}
where $M_x=m+n-2$, $M_y=m-n+N_2-1$, $q^x_{(A,B,1,2,3)}=(1/3,1,1/6,2/3,1/6)$, $q^y_{(A,B,1,2,3)}=(1/2,1/2,0,1/2,1)$, and  $\mathbf{P}=[P_x,P_y]$ are the gradients of the potential along the $x$ and $y$ directions; see \textbf{Figure}~\ref{fs4}.

Next, we show how to obtain the quasi-quantized displacement in the dynamics.
The evolution of the wavepacket $\varphi(z)$ initially prepared in the lowest Floquet band satisfies the equation, 
\begin{equation}
i\partial_z|\varphi(z)\rangle=[H_0(z)-H_p]|\varphi(z)\rangle.
\end{equation}
According to a well-established theory,\textsuperscript{\cite{xiao2010berry,ho2012quantized,zhu2021uncovering}} the group velocity of a wavepacket undergoing adiabatic evolution can be separated into two parts, the energy dispersion and Berry curvature of the individual energy bands, 
\begin{equation}
    v_{x(y)}(k_x,k_y,z)=\frac{\partial E(k_x,k_y,z)}{\partial k_{x(y)}}+\nu_{x(y)}(k_x,k_y,z).
\end{equation}
The first term leads to the conventional Bloch oscillations, while the second term results in a displacement perpendicular to the direction of the applied external force $\mathcal{F}_{x(y)}=P_{x(y)}/a_{x(y)}$.\textsuperscript{\cite{aidelsburger2015measuring}} 
The Bloch periods along $x$ and $y$ directions are given by $T_x=2\pi/(\mathcal F_x a_x)$ and $T_y=2\pi/(\mathcal F_y a_y)$, respectively.
To make sure the period of Bloch oscillations is the multiple of the driving period, we impose the ratio $T_o^x/T_o^y=\eta_x/\eta_y$ with $\eta_x$ and $\eta_y$ being coprime numbers, and furthermore, $T_o^x$ and $T_o^y$ are multiples of the driving period $T$.
Then, the overall Bloch period $T_o$ is the general multiplication of the driving period $T_o=\eta_yT_o^x=\eta_xT_o^y=nT$
The average position $ X(z)$ and $Y(z)$ of the wavepacket in the $x $ and $y $ directions can be calculated by
\begin{eqnarray}
        X(z)&=&\frac{\int \int  x|\varphi(x,y,z)|^2dxdy}{\int \int  |\varphi(x,y,z)|^2dxdy}, \\
         Y(z)&=&\frac{\int \int  y |\varphi(x,y,z)|^2dxdy}{\int \int  |\varphi(x,y,z)|^2dxdy},
\end{eqnarray}
where the $|\varphi(x,y,z)|^2$ is the probability density of wavepacket.
The mean displacement $\Delta X$ and $\Delta Y$ is given by
\begin{equation}\label{eqs11}
\begin{aligned}
        &\Delta X(z)=X(z)-X(0)=\int _0^z \frac{\partial E(k_x,k_y,z')}{\partial k_{x}}+\int_0^z \nu_{x}(k_x,k_y,z')dz',\\
        &\Delta Y(z)=Y(z)-Y(0)=\int _0^z \frac{\partial E(k_x,k_y,z')}{\partial k_{y}}+\int_0^z \nu_{y}(k_x,k_y,z')dz'.
\end{aligned}
\end{equation}
Because of the periodicity of Floquet Bloch bands, the integral of the dispersion velocity is exactly zero in the overall period. 
If the Bloch band is nontrivial, the anomalous group velocity thus plays a determinant role in the mean displacement. 
Similarly to the Reference~\cite{zhu2021uncovering}, we can define quasi-quantized numbers along the $x$ and $y$ directions as 
\begin{equation}
\begin{aligned}
    C_{x}&=-\frac{1}{2a_{x}\eta_x}\int _0^{T_o}\nu_{x}(k_x,k_y,z')dz',\\
    C_{y}&=\frac{1}{2a_{y}\eta_y}\int _0^{T_o}\nu_{y}(k_x,k_y,z')dz',
\end{aligned}
\end{equation}
which turn out to be nearly perfect quantized values due to uniform sweeping of the Brillouin zone.
Then, the displacement in an overall period tends to be quasi-quantized values,
\begin{eqnarray}
\begin{aligned}
&\Delta X(k_x,k_y,T_o)=-2a_x \eta_x C_x,\\
&\Delta Y(k_x,k_y,T_o)=2a_y\eta_y C_y.
\end{aligned}
\end{eqnarray}
Based on the above relations, the quasi-quantized numbers and the Chern number of a given Floquet band ($C$) are close to each other, $C_x\approx C_y\approx C$.

\medskip

%
\bibliographystyle{MSP}


\end{document}